\documentclass[%
 reprint,
 amsmath,amssymb,
 aps,
]{revtex4-2}

\usepackage{graphicx}
\usepackage{dcolumn}
\usepackage{bm}
\usepackage{color}
\usepackage{graphicx}                           
\usepackage{float}
\usepackage{amsmath}	
\usepackage{multirow} 
\usepackage{amssymb}	
\usepackage{diagbox}

\usepackage{natbib}

\begin{document}


\title{Cosmological Constraints from Bias-Robust Wavelet Scattering Statistics for Stage-IV Galaxy Surveys}

\author{Zhujun Jiang}
\affiliation{School of Physics and Astronomy, Sun Yat-sen University Zhuhai Campus, Zhuhai 519082, P.R.China.}
\author{Xu Xiao}%
\affiliation{School of Physics and Astronomy, Sun Yat-sen University Zhuhai Campus, Zhuhai 519082, P.R.China.}
\author{Zhiwei Min}%
\affiliation{School of Physics and Astronomy, Sun Yat-sen University Zhuhai Campus, Zhuhai 519082, P.R.China.}

\author{Zhao Chen}
\affiliation{Tsung-Dao Lee Institute, Shanghai Jiao Tong University, Shanghai 200240, China}
\affiliation{Department of Astronomy, School of Physics and Astronomy, Shanghai Jiao Tong University, Shanghai 200240, China}
\affiliation{State Key Laboratory of Dark Matter Physics, School of Physics and Astronomy, Shanghai Jiao Tong University, Shanghai 200240, China}
\affiliation{Key Laboratory for Particle Astrophysics and Cosmology (MOE)/Shanghai Key Laboratory for Particle Physics and Cosmology, Shanghai 200240, China}

\author{Yu Yu}
\affiliation{Department of Astronomy, School of Physics and Astronomy, Shanghai Jiao Tong University, Shanghai 200240, China}
\affiliation{State Key Laboratory of Dark Matter Physics, School of Physics and Astronomy, Shanghai Jiao Tong University, Shanghai 200240, China}
\affiliation{Key Laboratory for Particle Astrophysics and Cosmology (MOE)/Shanghai Key Laboratory for Particle Physics and Cosmology, Shanghai 200240, China}

\author{Fenfen Yin}
\email{dsjyff@gztrc.edu.cn}
\affiliation{Department of Physics and Electronic Engineering, Tongren University, Tongren 554300, China}

\author{Xiao-Dong Li}
\email{lixiaod25@mail.sysu.edu.cn}
\affiliation{School of Physics and Astronomy, Sun Yat-sen University Zhuhai Campus, Zhuhai 519082, P.R.China.\\
CSST Science Center for the Guangdong-Hong Kong-Macau Greater Bay Area, SYSU, Zhuhai 519082, P.R.China}

\author{Le Zhang}
\email{zhangle7@mail.sysu.edu.cn}
\affiliation{School of Physics and Astronomy, Sun Yat-sen University Zhuhai Campus, Zhuhai 519082, P.R.China.\\
CSST Science Center for the Guangdong-Hong Kong-Macau Greater Bay Area, SYSU, Zhuhai 519082, P.R.China
}

\date{\today}

\begin{abstract}
A central challenge in precision cosmology with galaxy surveys is to extract non-Gaussian information from large-scale structure while controlling systematic uncertainties such as tracer bias. Conventional clustering statistics, such as the two-point correlation function (2PCF), capture limited nonlinear information and typically require explicit bias modeling, which can introduce systematic errors if the adopted bias prescription is inaccurate. To address this problem, we introduce $R^{\rm wst}$, a bias-robust statistic constructed from $m$-mode ratios of the wavelet scattering transform (WST). Using simulation-based inference, we train a Gaussian-process-regression emulator on the \texttt{Kun} simulation suite and use \texttt{JiuTian} simulations for covariance estimation and validation. The emulator achieves percent-level accuracy, sufficient for the expected observational uncertainties. We show that $R^{\rm wst}$ yields unbiased constraints on $\Omega_m$, $\sigma_8$, $n_s$, and $w_0$, and improves the breaking of the $\Omega_m$--$\sigma_8$ degeneracy by about a factor of two compared with 2PCF. Its constraining power remains stable across a broad range of tracer-bias scenarios, demonstrating that $R^{\rm wst}$ can mitigate bias-induced systematics without explicit bias modeling. These results establish $R^{\rm wst}$ as a powerful and robust statistic for precision cosmology with Stage-IV surveys.
\end{abstract}


\maketitle{}

\section{Introduction}\label{sec:intro}

The large-scale structure (LSS) of the Universe encodes the expansion history of the cosmos and the growth of matter perturbations, and therefore provides one of the most powerful observational probes of dark energy, dark matter, and the formation of cosmic structure~\cite{2013PhR...530...87W}. Galaxies, as observable tracers of the underlying matter density field, have played a central role in this program. Over the past two decades, Stage-III galaxy surveys~\cite{2df:Colless:2003wz, beutler_6df_2012, beutler_clustering_2017, blake2011wigglez, blake2011wigglezb, york2000sloan, Eisenstein:2005su, Percival:2007yw, alam2017clustering, anderson2012clustering} have enabled precise measurements of baryon acoustic oscillations, redshift-space distortions, and cosmological parameters. The next generation of Stage-IV surveys, including DESI~\cite{levi_desi_2013}, LSST~\cite{collaboration_lsst_2009}, Euclid~\cite{laureijs_euclid_2011}, the Roman Space Telescope~\cite{eifler_cosmology_2021}, and the Chinese Space Station Telescope (CSST)~\cite{zhan_consideration_2011}, will provide galaxy samples with unprecedented volume, depth, and statistical power. These data sets create a major opportunity for precision cosmology, but also demand summary statistics and theoretical models that can fully exploit their nonlinear information content.

A central difficulty in extracting cosmological information from galaxy surveys is that galaxies are not unbiased tracers of the matter density field. Dark matter halos trace the matter distribution with a bias that depends on mass, environment, and scale~\cite{2010ApJ...724..878T}, while galaxies further trace halos through complex baryonic processes and halo occupation physics~\cite{Benson:1999mva}. On sufficiently large scales, this tracer bias can be modeled perturbatively~\cite{desjacques_large-scale_2018}. However, much of the statistical power of upcoming surveys lies in mildly nonlinear and nonlinear regimes, where the galaxy density field becomes strongly non-Gaussian and where bias modeling becomes increasingly uncertain. This creates a fundamental challenge: the most informative scales are also those on which conventional bias modeling is least robust.

Traditional two-point statistics, such as the power spectrum and the two-point correlation function (2PCF), remain indispensable tools for LSS cosmology, but they are not sufficient to capture the full information contained in nonlinear structure formation. Nonlinear gravitational evolution generates phase correlations and higher-order mode couplings that are invisible to Gaussian statistics. To recover this information, a variety of higher-order and non-Gaussian statistics have been explored, including $n$-point correlation functions~\cite{buchbinder_non-gaussianities_2008}, marked correlation functions~\cite{xiao_tomographic_2025}, Minkowski functionals~\cite{liu_cosmological_2025}, and density-split statistics~\cite{gon_turning_2025}. Deep-learning approaches, especially convolutional neural networks, have also demonstrated strong empirical performance in extracting non-Gaussian cosmological information~\cite{ribli_weak_2019}. Nevertheless, their limited interpretability and dependence on training choices make it difficult to identify which physical features drive the inferred constraints. An ideal statistic for Stage-IV LSS analyses should therefore combine sensitivity to nonlinear non-Gaussian information with interpretability and robustness to tracer bias.

The wavelet scattering transform (WST) provides a promising route toward this goal. Originally introduced as a stable and interpretable representation for high-dimensional signals~\cite{mallat_group_2012}, the WST constructs multiscale coefficients through cascades of wavelet convolutions and modulus operations, where the wavelet is powerful in data compression and de-noising~\cite{romeo_n-body_2003,romeo_wavelet_2004,romeo_discreteness_2008}. In contrast to fully trained neural networks, the WST is training-free and its coefficients have a clear connection to localized scale-dependent structures. At the same time, unlike two-point statistics, it is sensitive to non-Gaussian information generated by nonlinear mode coupling. These properties have motivated a growing range of applications in astrophysics and cosmology, including analyses of the 21 cm forest~\cite{shimabukuro_analyzing_2025}, EoR light-cone modeling~\cite{hothi_generative_2025}, 21 cm intensity mapping~\cite{greig_exploring_2022, greig_detecting_2023}, weak gravitational lensing~\cite{cheng_new_2020}, simulated three-dimensional density fields~\cite{valogiannis_towards_2022, jiang_new_2025}, BOSS CMASS galaxies~\cite{valogiannis_going_2022, blancard_galaxy_2024}, and primordial non-Gaussianity~\cite{peron_constraining_2024}.

Despite this progress, the use of WST statistics for galaxy clustering still faces the key obstacle of tracer bias. Previous studies have mitigated this effect by explicitly modeling galaxy bias through halo occupation distribution prescriptions~\cite{valogiannis_going_2022, valogiannis_towards_2022}. Such approaches are powerful, but they introduce additional modeling assumptions and nuisance parameters. In our previous work~\cite{jiang_new_2025}, we proposed a different strategy based on the WST $m$-mode ratio, denoted as $R^{\rm wst}$, together with high-density apodization. This statistic is designed to retain cosmological sensitivity while suppressing the dependence on tracer bias, thereby reducing the need for explicit bias modeling. If validated in a full inference framework, such a bias-robust non-Gaussian statistic would provide a valuable new observable for future galaxy surveys.

A second challenge is computational cost. Accurate inference with nonlinear non-Gaussian statistics requires predictions from high-resolution cosmological simulations across a multidimensional parameter space. Directly running such simulations inside a Bayesian inference pipeline is computationally infeasible. Emulators offer an efficient solution by learning a fast surrogate model from a finite suite of simulations. Cosmological emulators such as \texttt{CosmicEmu}~\cite{heitmann_coyote_2010, heitmann_coyote_2013}, \texttt{BACCO}~\cite{angulo_bacco_2021}, \texttt{EuclidEmulator}~\cite{euclid_collaboration_euclid_2019}, \texttt{AEMULUS}~\cite{zhai_aemulus_2019}, and \texttt{CSSTEmulator}~\cite{chen_csst_2025} have been developed for a range of observables, including the nonlinear matter power spectrum, halo statistics, and increasingly higher-order summary statistics. For WST-based inference, however, it remains necessary to establish whether a sufficiently accurate emulator can be constructed for a bias-robust WST statistic and whether the resulting inference remains reliable under realistic covariance and validation tests.

In this work, we develop a simulation-based inference framework for $R^{\rm wst}$. Using the \texttt{Kun}~\cite{chen_csst_2025} simulation suite, we train a Gaussian-process-regression (GPR) emulator for $R^{\rm wst}$ over the relevant cosmological parameter space. We estimate and validate the covariance using the \texttt{JiuTian} simulations, and perform Markov chain Monte Carlo analyses to quantify the cosmological constraining power of the statistic. Our main goals are threefold: 

1) to test whether $R^{\rm wst}$ can be emulated with sufficient accuracy for observational applications; 

2) to assess whether it provides competitive cosmological constraints beyond traditional two-point information; 

3) to examine whether its robustness to tracer bias persists in a full parameter-inference setting. This work therefore provides a critical step toward applying interpretable, bias-robust, non-Gaussian WST statistics to future Stage-IV galaxy surveys such as CSST.

The paper is organized as follows. In Section~\ref{sec:wst_m_mode_ratio}, we briefly review the WST $m$-mode ratio statistic, and describes the simulation data and preprocessing procedures. In Section~\ref{sec:emulator}, we introduce the construction and validation of the Gaussian-process emulator, and  presents the covariance matrix estimation and likelihood framework. The cosmological inference results and tracer-bias robustness tests are discussed in Section~\ref{sec:results}. Finally, we summarize our conclusions in Section~\ref{sec:conclusion}.

\section{The WST-based m-mode ratio statistic}
\label{sec:wst_m_mode_ratio}

\subsection{Three-dimensional wavelet scattering transform}
\label{subsec:basic_wst}

We first introduce WST, which provides the statistical basis for the WST-based $m$-mode ratio used in this work. The WST was originally developed as a non-learnable representation for high-dimensional signals. It applies a cascade of wavelet convolutions, nonlinear modulus operations, and spatial averaging. Because its filters are analytically defined rather than learned from data, the WST provides an interpretable and stable way to characterize complex fields. In particular, it is sensitive to non-Gaussian and higher-order statistical information that is not fully captured by two-point statistics.

For a three-dimensional input field $I(\bm{x})$, the WST probes the field with a family of localized wavelets. These wavelets are constructed at different spatial scales and angular modes, allowing the transform to extract structures with different characteristic sizes and directional dependence. After each wavelet convolution, a modulus operation is applied. This nonlinear step transfers information from small-scale fluctuations to larger-scale envelopes, making it possible to quantify correlations between structures across different scales and angular modes. The final spatial average produces translation-invariant WST coefficients.

In this work, we follow the 3D WST implementation introduced by \cite{jiang_new_2025} and compute scattering transform coefficients up to second order, using the open-source \textsc{Kymatio} package \citep{eickenberg_solid_2018}. We adopt a solid harmonic wavelet multiplied by a Gaussian envelope as the mother wavelet:
\begin{equation}
\Psi_l^m(\bm{x}) =
\frac{1}{(2\pi)^{3/2}}
\exp\left(-\frac{|\bm{x}|^2}{2\sigma^2}\right)
|\bm{x}|^l
Y_l^m\left(\frac{\bm{x}}{|\bm{x}|}\right)\,,
\end{equation}
where $Y_l^m$ denotes the spherical harmonic function, $l$ is the angular degree, $m$ is the azimuthal mode, and $\sigma$ controls the width of the Gaussian envelope. Throughout this study, we set $\sigma = 0.8$.

A dyadic family of wavelets is generated by dilating the mother wavelet:
\begin{equation}
\Psi_{j,l}^{m}(\bm{x})
=
2^{-3j}
\Psi_l^m(2^{-j}\bm{x})\,,
\end{equation}
where $j$ labels the spatial scale. Larger values of $j$ correspond to wavelets probing larger spatial scales.

We denote the input field as
\begin{equation}
U_0(\bm{x}) = I(\bm{x})\,.
\end{equation}
The first scattering operation convolves the input field with a wavelet and then applies the modulus:
\begin{equation}
U_1(j_1,l_1,m_1)(\bm{x})
=
\left|
U_0(\bm{x}) \otimes \Psi_{j_1,l_1}^{m_1}(\bm{x})
\right|\,,
\end{equation}
where $\otimes$ denotes convolution. The corresponding first-order WST coefficient is obtained by spatial averaging:
\begin{equation}
S_1(j_1,l_1,m_1)
=
\left\langle
\left|
U_0(\bm{x}) \otimes \Psi_{j_1,l_1}^{m_1}(\bm{x})
\right|^q
\right\rangle \,.
\end{equation}

The second-order WST coefficients are obtained by applying another wavelet convolution and modulus operation to the first-order modulus field:
{\small
\begin{equation}
S_2(j_2,l_2,m_2,j_1,l_1,m_1)
=\Big\langle\Big|U_1(j_1,l_1,m_1)(\bm{x}) \otimes
\Psi_{j_2,l_2}^{m_2}(\bm{x})
\Big|^q
\Big\rangle \,.
\end{equation}
}
These second-order coefficients measure how structures selected by the first wavelet are modulated by structures at another scale and angular mode. They therefore encode scale-scale and mode-mode couplings in the input field.

For completeness, the zeroth-order coefficient is defined as the spatial average of the input field:
\begin{equation}
S_0
=
\left\langle
|U_0(\bm{x})|^q
\right\rangle \,.
\end{equation}
The exponent $q$ controls which regions of the field are emphasized. Values $q > 1$ give more weight to high-amplitude regions, while values $q < 1$ enhance the relative contribution of low-amplitude regions. The standard WST corresponds to $q = 1$, which is the case adopted throughout this work.

Given a set of dyadic spatial scales $J$ and angular degrees $L$, the WST coefficients are indexed by
\begin{equation}
j \in [0,1,\dots,J]\,,~~l \in [0,1,\dots,L]\,,~~m \in [-l,\dots,l]\,.
\end{equation}
For second-order coefficients, we only keep scattering paths with $j_2 > j_1$, since paths with $j_2 < j_1$ have been shown to contain little additional cosmological information \citep{cheng_new_2020}. Retaining the explicit $m$ dependence of these coefficients allows us to construct the $m$-mode ratio statistic introduced in the following subsection.

\subsection{Construction of the WST $m$-mode ratio}
\label{subsec:wst_m_mode_ratio}

Based on the $m$-dependent WST coefficients introduced above, we construct the WST $m$-mode ratio statistic, which was first proposed by~\citep{jiang_new_2025} and was shown to be \textit{robust against tracer bias} while retaining sensitivity to cosmological parameters.

The main idea is to compare neighboring azimuthal modes of the WST coefficients at fixed spatial and angular scales. Since tracer bias mainly changes the overall amplitude of the field, taking ratios between adjacent $m$ modes can largely cancel such global amplitude modulation. At the same time, the relative variation among different $m$ modes preserves directional information, which is sensitive to anisotropic features in the underlying cosmological field.

For the first-order WST coefficients, the $m$-mode ratio is defined as
\begin{equation}
\label{eq:R1}
R_1^{\mathrm{wst}}(j_1,l_1,m_1)
=\frac{S_1(j_1,l_1,m_1+1)
}{S_1(j_1,l_1,m_1)
}\,,~l_1 \neq 0 \,.
\end{equation}

Similarly, for the second-order WST coefficients, we define
\small{
\begin{equation}
\label{eq:R2}
R_2^{\mathrm{wst}}(j_2,j_1,l_1,m_1)
=\frac{S_2(j_2,j_1,l_1,m_1+1)
}{S_2(j_2,j_1,l_1,m_1)
}\,,~l_1 \neq 0 \,.
\end{equation}
}
Here, $S_1$ and $S_2$ denote the first- and second-order WST coefficients, respectively. For a real-valued input field, the WST coefficients satisfy the symmetry relation
\begin{equation}
S^m = S^{-m}.
\end{equation}
Therefore, we only use the coefficients with $m \geq 0$, since the negative-$m$ modes do not provide independent information. The case $l = 0$ is excluded because it contains only the $m = 0$ mode, and no ratio between neighboring $m$ modes can be constructed.

For the purpose of this study, we extend the original WST ratio statistic by explicitly retaining the $m$-mode dependence, rather than averaging over $m$ as in \citep{jiang_new_2025}. This extension preserves additional directional information encoded in the relative amplitudes of different azimuthal harmonic modes. We find that the resulting $m$-dependent ratios yield for the slightly improved cosmological constraints compared to the original $m$-averaged statistic.

In this study, we use $j_1 \in [0,5]$, and $l_1 \in [0,3]$ for the first-order WST coefficients. For the second-order coefficients, we fix $j_1 = 3$, $j_2 = 4,5$ and $l_1 \in [0,3]$. Only modes with $l_1 \neq 0$ are included when constructing the ratio statistic. With this configuration, the final WST $m$-mode ratio data vector contains the selected $R_1^{\mathrm{wst}}$ and $R_2^{\mathrm{wst}}$ coefficients used in our cosmological analysis.

Physically, this ratio statistic measures how the WST response changes across neighboring azimuthal modes. Since it compares relative amplitudes rather than absolute coefficient values, it is less sensitive to isotropic amplitude changes induced by tracer bias. Meanwhile, it remains sensitive to anisotropic and non-Gaussian structures in the field. The mathematical motivation and numerical validation of this statistic are discussed in detail in~\citep{jiang_new_2025}.

\subsection{The \texttt{Kun} and \texttt{JiuTian} simulations}
\label{subsec:jiutian_simulations}

We use halo catalogs from the \texttt{Kun} cosmological simulation suite \citep{chen_csst_2025}, which is part of the broader \texttt{JiuTian} simulation program \citep{han_jiutian_2025,chen_cosmological_2025,yu_cube2_2026,he_extending_2023,zhang_first_2019,luo_elucid_2024,pei_simulating_2024,gu_csst_2024,tan_semi-analytical_2025,wei_mock_2026,wei_mock_2026a}. The \texttt{Kun} suite was designed to support the upcoming CSST survey and covers a broad range of cosmological models within the $w_0w_a{\rm CDM}+\sum m_\nu$ framework. In this cosmological model, CDM stands for cold dark matter. The dark energy equation of state is described by two parameters, $(w_0, w_a)$, where $w_0$ is its present-day value and $w_a$ describes how it changes with time. The term $\sum m_\nu$ represents the total neutrino mass.

The \texttt{Kun} simulations sample an eight-dimensional cosmological parameter space consisting of
\begin{equation}
\left\{
\Omega_b,\,
\Omega_{cb},\,
H_0,\,
n_s,\,
A_s,\,
w_0,\,
w_a,\,
\sum m_\nu
\right\}.
\end{equation}
Here, $\Omega_b$ is the present-day baryon density parameter, $\Omega_{cb}$ is the combined density parameter of cold dark matter and baryons, $H_0$ is the Hubble constant, $n_s$ is the scalar spectral index, and $A_s$ is the amplitude of the primordial scalar power spectrum. 

We use a total of 129 simulations, including 128 non-fiducial cosmologies sampled by Sobol sequences~\cite{chen_csst_2025} over the parameter ranges $\Omega_b \in [0.04,0.06]$, $\Omega_{cb} \in [0.24,0.40]$, $n_s \in [0.92,1.00]$, $H_0 \in [60,80]~{\rm km}~{\rm s}^{-1}~{\rm Mpc}^{-1}$, $A_s \in [1.7,2.5]\times10^{-9}$, $w_0 \in [-1.3,-0.7]$, $w_a \in [-0.5,0.5]$, and $\sum m_\nu \in [0.00,0.30]~{\rm eV}$, together with one fiducial simulation based on the {\it Planck} 2018 cosmology \citep{planck_collaboration_planck_2020}.

All \texttt{Kun} simulations were evolved with the \texttt{Gadget-4} $N$-body solver. Each simulation contains $3072^3$ particles in a periodic comoving box with side length $1~h^{-1}~{\rm Gpc}$, where $h \equiv H_0/(100~{\rm km}~{\rm s}^{-1}~{\rm Mpc}^{-1})$. This mass and force resolution enables accurate modeling of matter clustering into the deeply nonlinear regime, up to approximately $k \lesssim 10~h~{\rm Mpc}^{-1}$. The initial density fields are generated using the fixed-amplitude method, which suppresses sample variance among simulations.

The simulations are initialized at redshift $z=127$ and evolved to $z=0$, with 12 snapshots saved between $z=3$ and $z=0$ using an approximately linear redshift step size of $\Delta z \sim 0.25$. At each snapshot, dark matter halos and subhalos are identified using both \texttt{SubFind} \citep{springel_populating_2001} and \texttt{Rockstar} \citep{behroozi_rockstar_2013}. 

In this work, we focus on the halo catalogs at $z=0.5$. This choice is motivated by upcoming CSST-like Stage-IV slitless spectroscopic galaxy surveys, which are expected to observe a large number of galaxies primarily at redshifts $z < 1$. We use the 129 \texttt{Kun} simulations, labeled \texttt{c0000--c0128}, to construct the training and validation data for our emulator. The simulation  \texttt{c0000} corresponds to the fiducial {\it Planck} 2018 cosmology.

For covariance matrix estimation, we use the large-volume \texttt{JiuTian} simulation from the \texttt{JiuTian} suite \citep{han_jiutian_2025}. This simulation follows the {\it Planck} 2018 cosmology (the same cosmology as \texttt{c0000} in \texttt{Kun}) and contains $6144^3$ particles in a periodic box with side length $2~h^{-1}~{\rm Gpc}$. Its larger volume provides a more suitable dataset for estimating the covariance of the WST $m$-mode ratio statistics.

\subsection{Construction of analysis datasets}
\label{subsec:analysis_datasets}

We construct the analysis datasets from the halo catalogs described in Section~\ref{subsec:jiutian_simulations}. These datasets are used for two purposes: first, to train emulators for the WST $m$-mode ratio statistics, and second, to test the robustness of these statistics against variations in tracer bias.

Since halo bias is mainly determined by halo mass at leading order, we generate halo samples with different mass selections to mimic tracers with different bias levels. The emulator training set is constructed from the \texttt{Kun} simulations, while the mock observational data vectors and covariance matrix are constructed from the large-volume \texttt{JiuTian} simulation.

For emulator training, we select dark matter halos from each of the 129 \texttt{Kun} simulations using a fixed number density of $n = 10^{-3}~h^3~{\rm Mpc}^{-3}$. Because the halo mass function depends on cosmology, the corresponding halo mass threshold varies among the 129 \texttt{Kun} cosmologies.

For covariance estimation and robustness tests, we construct three halo samples from the \texttt{JiuTian} simulation. These samples have the same number density, $n = 10^{-3}~h^3~{\rm Mpc}^{-3}$, but different halo mass ranges. By shifting the selected mass range toward lower masses, we obtain mock tracers with different effective bias levels:
\begin{enumerate}
    \item $M^A_{\rm cut}$: the fiducial halo-mass selection, matched to the \texttt{Kun} training sample and covariance estimation, with $M_{\rm min} = 3.86 \times 10^{12}~h^{-1}M_\odot$ and $M_{\rm max} = 1.93 \times 10^{15}~h^{-1}M_\odot$. This selection corresponds to the highest halo-mass range among the three cases considered.

    \item $M^B_{\rm cut}$: shift toward lower halo masses by $1.17\%$ relative to $M^A_{\rm cut}$, with $M_{\rm min} = 3.82 \times 10^{12}~h^{-1}M_\odot$ and $M_{\rm max} = 8.90 \times 10^{14}~h^{-1}M_\odot$.

    \item $M^C_{\rm cut}$: shift toward lower halo masses by $3\%$ relative to $M^A_{\rm cut}$, with $M_{\rm min} = 3.76 \times 10^{12}~h^{-1}M_\odot$ and $M_{\rm max} = 5.30 \times 10^{13}~h^{-1}M_\odot$.
\end{enumerate}

\begin{figure}
\centering
\includegraphics[scale=0.42]{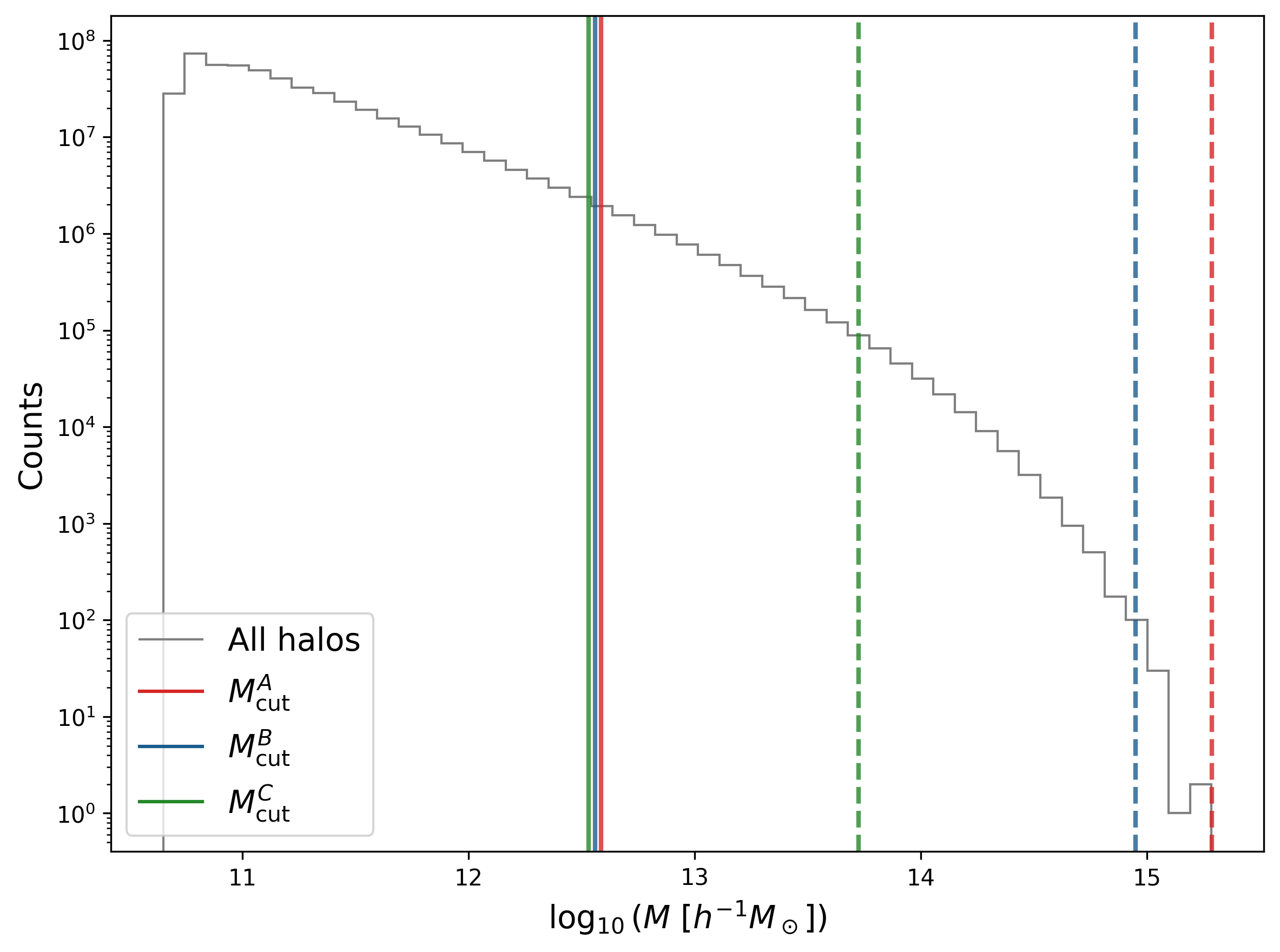}
\caption{
Mass distributions of the three mass-selected halo samples, $M^A_{\rm cut}$, $M^B_{\rm cut}$, and $M^C_{\rm cut}$, with the same number density $n = 10^{-3}~h^3 \rm{Mpc}^{-3}$, compared with the full halo catalog. Since the $M_{\rm min}$ values are very close across the three samples, we slightly offset them in the figure for clarity.
}
\label{fig:mass_function}
\end{figure}

Figure~\ref{fig:mass_function} shows the mass distributions of the three selected halo samples compared with the full halo catalog. Relative to the fiducial $M^A_{\rm cut}$ sample, $M^B_{\rm cut}$ has a moderately reduced high-mass tail, while $M^C_{\rm cut}$ strongly suppresses the contribution from massive halos. These three samples therefore span a range from mild to extreme tracer-bias variations, allowing us to test whether the WST $m$-mode ratio statistic remains stable under changes in halo bias.

The emulator is trained only on the \texttt{Kun} simulations with the $M^A_{\rm cut}$ selection. Once trained, the emulator is kept fixed in all subsequent parameter-inference analyses. The covariance matrix is estimated from subvolumes of the \texttt{JiuTian} simulation using the $M^A_{\rm cut}$ selection. For each of the three halo-mass selections, we construct a mock observational data vector from the corresponding \texttt{JiuTian} halo sample and perform MCMC inference using the same emulator and the same covariance matrix. This setup isolates the impact of tracer-bias variations on the inferred cosmological parameters.

We then convert the selected halo catalogs into three-dimensional density fields. Halos are assigned to uniform grids using the Cloud-in-Cell (CIC) interpolation scheme. We adopt a fixed cell size of $\Delta x = 5~h^{-1}~{\rm Mpc}$, which gives a grid size of $200^3$ for the \texttt{Kun} simulations and $400^3$ for the \texttt{JiuTian} simulation, corresponding to their box side lengths of $1~h^{-1}~{\rm Gpc}$ and $2~h^{-1}~{\rm Gpc}$, respectively.

Before computing the WST statistics, we apply the high-density apodization scheme introduced by \citet{jiang_new_2025}. This preprocessing step suppresses extremely high-density regions, where the WST coefficients can be overly sensitive to sharp density gradients. The apodized density field is defined as
\begin{equation}
\label{eq:high_density_apodization}
\rho_{\rm apo}(\bm{x})
=
\frac{\rho(\bm{x})}{2}
\left[1+\tanh(\alpha)\right]
+
\frac{\bar{\rho}}{2}
\left[1-\tanh(\alpha)\right],
\end{equation}
where $\alpha = k[\rho_\epsilon-\rho(\bm{x})]$, $\rho(\bm{x})$ is the original density field, $\rho_{\rm apo}(\bm{x})$ is the apodized density field, $\bar{\rho}$ is the global mean density, $\rho_\epsilon$ is the density threshold above which apodization becomes effective, and $k$ controls the sharpness of the transition.

This prescription leaves low- and moderate-density regions nearly unchanged, while smoothly replacing very high-density regions by values closer to the mean density. In this way, the apodization reduces the disproportionate influence of rare, massive structures on the WST coefficients.

For each density field, we sort all grid-cell density values in descending order and define $\rho_\epsilon$ as the density value corresponding to the top $4\%$ of cells. For the three \texttt{JiuTian} halo-mass selections, the thresholds are $\rho_\epsilon = 2.07 \times 10^{13}$, $1.57 \times 10^{13}$, and $1.36 \times 10^{13}$ $(h^{-1} {\rm Mpc})^{-3}$ for $M^A_{\rm cut}$, $M^B_{\rm cut}$, and $M^C_{\rm cut}$, respectively. We set $k = 0.4$ throughout this work.

After apodization, the density field is converted into the overdensity field $\delta(\bm{x}) = [\rho_{\rm apo}(\bm{x})-\bar{\rho}_{\rm apo}]/\bar{\rho}_{\rm apo}$, where $\bar{\rho}_{\rm apo}$ is the mean of the apodized density field. The resulting overdensity fields are then used to compute the WST $m$-mode ratio statistics. This preprocessing procedure is specifically designed for the WST-based statistics used in this work. It reduces the sensitivity of the WST coefficients to rare high-density peaks while preserving the cosmological information encoded in the large-scale and intermediate-scale structure of the halo density field.

\section{Gaussian-process emulator construction and validation}
\label{sec:emulator}

To perform Bayesian parameter inference efficiently, we construct emulators for the WST $m$-mode ratio statistics. An emulator is a surrogate model trained on a finite set of simulations. It learns the mapping from cosmological parameters to summary statistics and can then rapidly predict these statistics at new cosmologies within the training parameter space. In our case, the emulator approximates the dependence of each component of the data vector $R^{\rm wst}$ on the cosmological parameters of the $w_0w_a{\rm CDM}+\sum m_\nu$ model.

Emulation is particularly useful here because running new high-resolution $N$-body simulations for every point sampled in an MCMC analysis would be computationally prohibitive. Instead, the emulator provides fast interpolated predictions based on the 129 \texttt{Kun} simulations. Recent developments in machine-learning-based emulation have made this approach widely used in cosmological analyses~\citep{valogiannis_precise_2024}.

In this work, we use Gaussian Process Regression (GPR) to construct the emulator. GPR is well suited to our problem for two main reasons. First, it performs robustly when the number of training simulations is limited. Second, it is probabilistic, meaning that it provides both a predicted mean and an associated interpolation uncertainty. This allows the emulator uncertainty to be quantified and, if necessary, propagated into the likelihood analysis.

Following the general strategy of the CSST Emulator~\citep{chen_csst_2025}, we train the emulator on the \texttt{Kun} simulations to model the cosmological dependence of the summary statistics. However, our implementation differs from the CSST Emulator in the treatment of the output data vector. The CSST Emulator first applies Principal Component Analysis (PCA) to reduce the dimensionality of high-dimensional summary statistics before applying GPR. In contrast, the WST $m$-mode ratio data vector used in this work has a moderate dimension of $n_R=48$. Therefore, we do not apply PCA. Instead, we train an independent GPR model for each component of $R^{\rm wst}$.

More explicitly, for each statistic component $R_i^{\rm wst}$, where $i=1,\dots,n_R$, we train a separate GPR emulator as a function of the 8 cosmological parameters, including $\Omega_b$, $\Omega_{m}$, $n_s$, $A_s$, $w_0$, $w_a$, $\sigma_8$ and $H_0$. The full emulator prediction is then obtained by combining the predictions from all 48 independent GPR models. This direct component-wise emulation avoids information loss from dimensionality reduction and keeps the interpretation of each WST $m$-mode ratio component transparent.

\subsection{Gaussian process regression}
\label{subsec:gpr}

We use Gaussian Process Regression (GPR) to emulate the dependence of each WST $m$-mode ratio component on cosmological parameters. For each component of the data vector, denoted by $R_i^{\rm wst}$, we train an independent GPR model. The input of the emulator is the eight-dimensional cosmological parameter vector $\boldsymbol{\theta}$, and the output is the corresponding statistic component $R_i^{\rm wst}(\boldsymbol{\theta})$.

A Gaussian process defines a probability distribution over functions. More specifically, any finite set of function values drawn from a Gaussian process follows a joint Gaussian distribution \citep{rasmussen_gaussian_2008}. For one statistic component, we model the unknown function as $f(\boldsymbol{\theta}) \sim \mathcal{GP}[m(\boldsymbol{\theta}), k(\boldsymbol{\theta},\boldsymbol{\theta}')]$, where $m(\boldsymbol{\theta})$ is the mean function and $k(\boldsymbol{\theta},\boldsymbol{\theta}')$ is the covariance function, or kernel. The mean function describes the prior expectation of the statistic at a given point in parameter space, while the kernel determines how strongly two function values are correlated as a function of their separation in cosmological parameter space.

Before training the emulator, we normalize the input cosmological parameters to improve numerical stability and to avoid parameters with larger numerical ranges dominating the kernel distance. We adopt a zero mean function, $m(\boldsymbol{\theta})=0$, which is a standard choice in GPR after centering the training targets.

For a given statistic component, the training set is written as $\mathcal{D}=\{(\boldsymbol{\theta}_a,y_a)\}_{a=1}^{N}$, where $N$ is the number of training cosmologies and $y_a=R_i^{\rm wst}(\boldsymbol{\theta}_a)$. We model the training targets as $y_a=f(\boldsymbol{\theta}_a)+\epsilon_a$, with $\epsilon_a \sim \mathcal{N}(0,\sigma_n^2)$. Here, $\sigma_n$ acts as an effective noise or nugget term, which improves numerical stability and accounts for small residual uncertainties in the training data.

Let $\boldsymbol{\Theta}=\{\boldsymbol{\theta}_1,\dots,\boldsymbol{\theta}_N\}$ be the full set of training inputs, $\mathbf{y}=(y_1,\dots,y_N)^T$ be the training target vector, and $\boldsymbol{\theta}_*$ be a test cosmology. The joint prior distribution of the training targets and the latent function value $f_*=f(\boldsymbol{\theta}_*)$ is
\begin{equation}
\begin{bmatrix}
\mathbf{y} \\
f_*
\end{bmatrix}
\sim
\mathcal{N}
\left(
\mathbf{0},
\begin{bmatrix}
\mathbf{K}+\sigma_n^2\mathbf{I} & \mathbf{k}_* \\
\mathbf{k}_*^T & k_{**}
\end{bmatrix}
\right),
\end{equation}
where $\mathbf{K}=k(\boldsymbol{\Theta},\boldsymbol{\Theta})$ is the covariance matrix of the training inputs, $\mathbf{k}_*=k(\boldsymbol{\Theta},\boldsymbol{\theta}_*)$ is the covariance vector between the training inputs and the test point, and $k_{**}=k(\boldsymbol{\theta}_*,\boldsymbol{\theta}_*)$ is the prior variance at the test point.

Conditioning this joint Gaussian distribution on the training data gives the posterior predictive distribution. The predicted mean and variance at $\boldsymbol{\theta}_*$ are
\begin{align}
\bar{f}_*
&=
\mathbf{k}_*^T
\left(
\mathbf{K}+\sigma_n^2\mathbf{I}
\right)^{-1}
\mathbf{y},
\\
{\rm Var}(f_*)
&=
k_{**}
-
\mathbf{k}_*^T
\left(
\mathbf{K}+\sigma_n^2\mathbf{I}
\right)^{-1}
\mathbf{k}_* .
\end{align}
The predictive mean $\bar{f}_*$ is used as the emulator prediction for the statistic component, while ${\rm Var}(f_*)$ quantifies the interpolation uncertainty of the emulator.

We adopt a constant kernel multiplied by an anisotropic radial basis function (RBF) kernel. This choice assumes that cosmologies closer to each other in the normalized parameter space should give more strongly correlated WST $m$-mode ratio statistics. The kernel is written as
\begin{equation}
k(\boldsymbol{\theta}_i,\boldsymbol{\theta}_j)
=
C
\exp
\left[
-\frac{1}{2}
\sum_{p=1}^{d}
\frac{
(\theta_{i,p}-\theta_{j,p})^2
}{
l_p^2
}
\right],
\end{equation}
where $d=8$ is the dimension of the cosmological parameter space, $C$ controls the overall amplitude of the covariance, and $l_p$ is the characteristic length scale along the $p$-th cosmological parameter direction. A smaller $l_p$ means that the statistic varies more rapidly with that parameter, while a larger $l_p$ corresponds to a smoother dependence.

The kernel hyperparameters, including $C$, the set of length scales $\{l_p\}_{p=1}^{d}$, and the nugget parameter $\sigma_n$, are determined by maximizing the log marginal likelihood of the training data:
\begin{equation}
\begin{aligned}
\ln \mathcal{L}_{\rm GPR}
=&
-\frac{1}{2}
\mathbf{y}^T
\left(
\mathbf{K}+\sigma_n^2\mathbf{I}
\right)^{-1}
\mathbf{y}
\\
&-\frac{1}{2}
\ln
\left|
\mathbf{K}+\sigma_n^2\mathbf{I}
\right|
-\frac{N}{2}\ln(2\pi).
\end{aligned}
\end{equation}
The first term measures how well the emulator fits the training data, the second term penalizes overly complex covariance structures, and the last term is a normalization constant. After optimization, the trained GPR model provides fast predictions of $R_i^{\rm wst}$ at any cosmology within the sampled parameter range.

\subsection{Emulator validation}
\label{subsec:emulator_validation}

\begin{figure*}[htpb]
\centering
\includegraphics[scale=0.6]{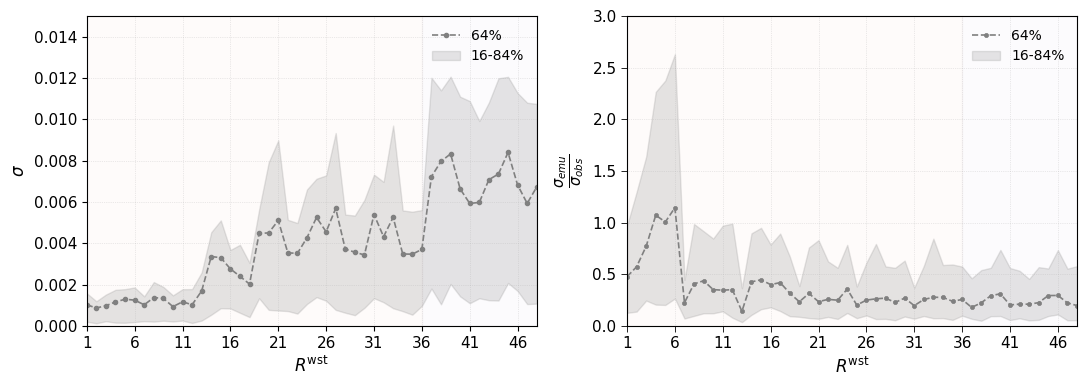}
\caption{
\label{fig:LOO_error&error_ratios}
Validation performance of the emulator for the 48 $R^{\rm wst}$ coefficients, including 36 first-order coefficients $R_1^{\rm wst}$ and 12 second-order coefficients $R_2^{\rm wst}$. The left panel shows the LOO relative errors for all coefficients. The right panel shows the ratio of the LOO absolute error to the effective observational uncertainty, where the latter is estimated from subvolume patch statistics of the \texttt{JiuTian} simulation with the $M^A_{\rm cut}$ selection. Different shaded colors indicate the $R_1^{\rm wst}$ and $R_2^{\rm wst}$ coefficients.
}
\end{figure*}

\begin{figure*}[htpb]
\centering
\includegraphics[scale=0.6]{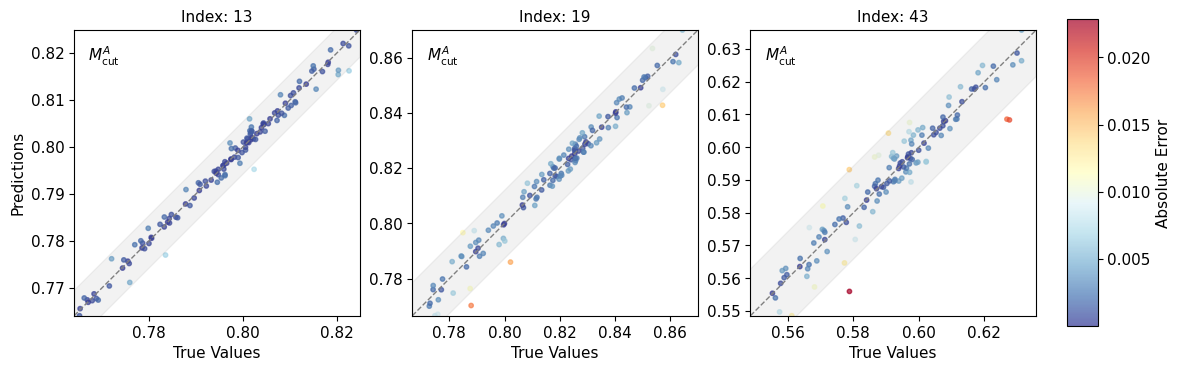}
\caption{
\label{fig:scatter_fig}
Comparison between the true and emulator-predicted values from the LOO tests for representative $R^{\rm wst}$ coefficients. The titles indicate the corresponding $(j,l,m)$ indices. The scatter points show the true and predicted values for all validation samples, and the color indicates the absolute prediction error. The gray band represents the $1\sigma$ effective observational uncertainty.
}
\end{figure*}

After constructing the emulator, we assess its predictive performance before applying it to MCMC parameter inference. To do so, we use leave-one-out (LOO) cross-validation. In this procedure, one simulation is removed from the training set and treated as a validation sample, while the emulator is trained on all remaining simulations. This process is repeated until every simulation has been used once as the validation sample. In this way, we obtain a prediction error for each cosmology in the training set and can evaluate the overall interpolation accuracy of the emulator.

For each $R^{\rm wst}$ component, we summarize the LOO performance using the 68th percentile of the error distribution over all validation samples, which we interpret as the effective $1\sigma$ emulator uncertainty. This provides a robust measure of the typical prediction error and avoids overemphasizing a small number of outliers.

The emulator performance for the $M^A_{\rm cut}$ selection is summarized in Fig.~\ref{fig:LOO_error&error_ratios} and Fig.~\ref{fig:scatter_fig}. In Fig.~\ref{fig:LOO_error&error_ratios}, the left panel shows that the LOO relative errors are below $1.2\%$ for all 48 $R^{\rm wst}$ components, spanning different scales and angular modes. This demonstrates that the emulator can accurately interpolate the cosmological dependence of the WST $m$-mode ratio statistics across the training set.

The right panel of Fig.~\ref{fig:LOO_error&error_ratios} places the emulator error in the context of the expected data uncertainty. Specifically, we compare the LOO absolute error, denoted by $\sigma_{\rm emu}$, with an effective observational uncertainty $\sigma_{\rm obs}$ estimated from subvolume patch statistics of the \texttt{JiuTian} simulation under the $M^A_{\rm cut}$ selection. We find that $\sigma_{\rm emu}/\sigma_{\rm obs}$ is below unity for almost all coefficients, with a mean value of $\langle \sigma_{\rm emu}/\sigma_{\rm obs} \rangle = 0.35$. This shows that the emulator error is generally much smaller than the statistical uncertainty expected in the data, and is therefore subdominant in the subsequent likelihood analysis.

We note that two coefficients have $\sigma_{\rm emu}/\sigma_{\rm obs} > 1$. These outliers correspond to modes with small $j$, $l$, and $m$ indices, i.e. a large-scale statistic. Such modes are more sensitive to cosmic variance and therefore exhibit a larger intrinsic scatter in $R^{\rm wst}$, making them more difficult to emulate precisely.

Figure~\ref{fig:scatter_fig} provides a complementary visualization of the emulator accuracy by directly comparing the predicted and true values from the LOO tests for several representative $R^{\rm wst}$ coefficients. The dashed line indicates the ideal one-to-one relation, $f_*^{\rm pred} = f_*^{\rm true}$. The points are tightly distributed around this line, indicating that the emulator predictions are in good agreement with the true values across the full range of training cosmologies. The scatter points do not show any obvious systematic offset, which suggests that the emulator is not biased.

Taken together, these validation tests show that the GPR emulator achieves high interpolation accuracy for both $R_1^{\rm wst}$ and $R_2^{\rm wst}$ coefficients. Its prediction error is small compared with the expected statistical uncertainty of the data, which confirms that the emulator is sufficiently accurate for the cosmological parameter inference presented in the following sections.

\subsection{Covariance matrix and likelihood}
\label{sec:covariance_likelihood}

After constructing and validating the emulator, we now describe the covariance matrix and likelihood framework used for cosmological parameter inference. The data vector used in the likelihood is denoted by $\mathbf{R}$. Depending on the analysis, $\mathbf{R}$ may contain the WST $m$-mode ratio statistics $R^{\rm wst}$ alone, or the normalized 2PCF, $\widehat{\xi}(s)$.

We assume that the data vector follows a multivariate Gaussian distribution. For a given cosmological parameter vector $\boldsymbol{\theta}$, the log-likelihood is written as
\begin{equation}
\label{eq:log_likelihood}
\ln \mathcal{L}(\boldsymbol{\theta}|\mathbf{d})
=
-\frac{1}{2}
\left[
\mathbf{R}_{\rm d}
-
\mathbf{R}_{\rm t}(\boldsymbol{\theta})
\right]^T
\mathbf{C}^{-1}
\left[
\mathbf{R}_{\rm d}
-
\mathbf{R}_{\rm t}(\boldsymbol{\theta})
\right],
\end{equation}
where $\mathbf{R}_{\rm d}$ is the data vector measured from the mock observational catalog, $\mathbf{R}_{\rm t}(\boldsymbol{\theta})$ is the theoretical prediction from the emulator, and $\mathbf{C}$ is the total covariance matrix.

The total covariance consists of two main contributions: the statistical uncertainty of the data and the emulator prediction error. We write this schematically as
\begin{equation}
\mathbf{C} = \mathbf{C}_{\rm d} + \mathbf{C}_{\rm emu},
\end{equation}
where $\mathbf{C}_{\rm d}$ represents the data covariance and $\mathbf{C}_{\rm emu}$ represents the emulator uncertainty.

The dominant contribution is the data covariance $\mathbf{C}_{\rm d}$, which is estimated from the large-volume \texttt{JiuTian} simulation with the $M^A_{\rm cut}$ selection. We divide the full $(2~h^{-1}{\rm Gpc})^3$ simulation box into $8^3=512$ subvolumes, each with side length $250~h^{-1}{\rm Mpc}$. The statistic vector is measured in each subvolume. The covariance estimated from these subvolumes is then rescaled to the target observational volume:
\begin{equation}
\label{eq:data_covariance}
\mathbf{C}_{\rm d}
=
\frac{V_{\rm sub}}{V_{\rm d}}
\frac{1}{N_{\rm sub}-1}
\sum_{k=1}^{N_{\rm sub}}
\left(
\mathbf{R}_k-\bar{\mathbf{R}}
\right)
\left(
\mathbf{R}_k-\bar{\mathbf{R}}
\right)^T .
\end{equation}
Here, $N_{\rm sub}=512$, $\mathbf{R}_k$ is the statistic vector measured from the $k$-th subvolume, $\bar{\mathbf{R}}$ is the mean vector over all subvolumes, $V_{\rm sub}$ is the subvolume size, and $V_{\rm d}$ is the effective volume of the mock observational data. The factor $V_{\rm sub}/V_{\rm d}$ rescales the covariance from the subvolume scale to the observational volume.

The emulator covariance $\mathbf{C}_{\rm emu}$ is estimated from the leave-one-out validation residuals described in Section~\ref{sec:emulator}. For each \texttt{Kun} cosmology in the LOO test, we compare the emulator prediction with the true statistic measured from the simulation. The emulator covariance is computed as
\begin{equation}
\label{eq:emulator_covariance}
\mathbf{C}_{\rm emu}
=
\frac{1}{N_{\rm mock}-1}
\sum_{j=1}^{N_{\rm mock}}
\left(
\mathbf{R}_{{\rm p},j}
-
\mathbf{R}_{{\rm t},j}
\right)
\left(
\mathbf{R}_{{\rm p},j}
-
\mathbf{R}_{{\rm t},j}
\right)^T ,
\end{equation}
where $\mathbf{R}_{{\rm p},j}$ is the emulator prediction for the $j$-th LOO test sample, $\mathbf{R}_{{\rm t},j}$ is the corresponding true statistic measured from the \texttt{Kun} simulation, and $N_{\rm mock}$ is the number of LOO validation samples. As shown in Section~\ref{sec:emulator}, the emulator uncertainty is generally subdominant compared with the expected observational uncertainty.

Following~\citep{yuan_stringent_2022}, we account for the fixed-phase effect in the emulator predictions. This effect arises because the emulator is trained on simulations with fixed initial phases, which suppresses sample variance in the training set but can introduce a small offset relative to the ensemble-mean statistic expected for random phases. To mitigate this, we apply a ratio-based correction instead of adding an explicit covariance term:

\begin{equation}
\label{eq:phase_correction}
\mathbf{R}_{\rm corr}(\boldsymbol{\theta})
=
\mathbf{R}_{\rm emu}(\boldsymbol{\theta})
\odot
\frac{\bar{\mathbf{R}}_{\rm JiuTian}}{\mathbf{R}_{\rm Kun,fid}}\,,
\end{equation}
where $\mathbf{R}_{\rm emu}(\boldsymbol{\theta})$ is the raw emulator prediction, $\bar{\mathbf{R}}_{\rm JiuTian}$ is the mean statistic measured from the \texttt{JiuTian} subvolumes at the fiducial cosmology, $\mathbf{R}_{\rm Kun,fid}$ is the statistic from the fiducial \texttt{Kun} simulation, and $\odot$ denotes element-wise multiplication. This correction assumes that the ratio between random-phase and fixed-phase measurements is approximately independent of cosmology. Previous work has shown that the ratio-based correction produces results comparable to explicitly including a phase covariance term in likelihood analyses \citep{valogiannis_precise_2024}. In the following inference, we use the corrected emulator prediction $\mathbf{R}_{\rm corr}(\boldsymbol{\theta})$ in place of $\mathbf{R}_{\rm t}(\boldsymbol{\theta})$.

\begin{figure}[htpb]
\centering
\includegraphics[scale=0.6]{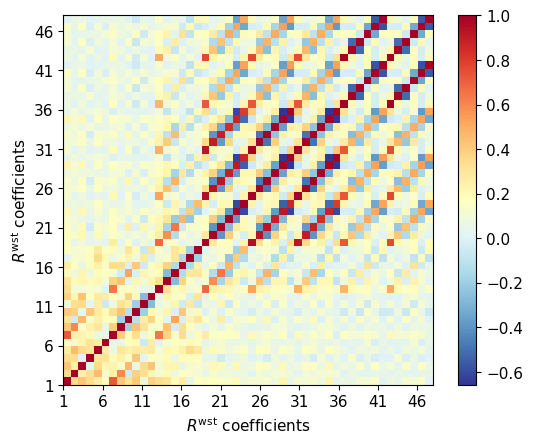}
\caption{
\label{fig:corre_obs_emu}
Correlation matrix of the 48 $R^{\rm wst}$ coefficients used in our analysis, estimated from the 512 subvolumes of the \texttt{JiuTian} simulation with the $M^A_{\rm cut}$ selection as an example. The coefficients are ordered by increasing values of their $j$, $l$, and $m$ indices, with the first 36 entries corresponding to $R_1^{\rm wst}$ and the last 12 entries corresponding to $R_2^{\rm wst}$.
}
\end{figure}

Figure~\ref{fig:corre_obs_emu} shows the correlation matrix of the 48-dimensional $R^{\rm wst}$ data vector estimated from the \texttt{JiuTian} simulation at the fiducial cosmology. The first 36 components correspond to first-order WST $m$-mode ratios, while the remaining 12 components correspond to second-order ratios. As expected, the diagonal elements are equal to unity. The off-diagonal elements show both positive and negative correlations among different coefficients.

The correlations are strongest between nearby components in the data vector. This behavior is expected because neighboring WST coefficients probe related wavelet scales and angular modes. In addition, the ratio definition of $R^{\rm wst}$ can introduce correlations between adjacent $m$ modes. Since the second-order coefficients are constructed using $j_1=3$ and $j_2=4,5$, they show mild correlations with first-order coefficients at nearby scales. This reflects the hierarchical structure of the WST, where second-order coefficients quantify how first-order wavelet responses are modulated on larger scales.

In contrast, widely separated components show weak, negligible, or mildly negative correlations. This indicates that different WST $m$-mode ratio components capture complementary information from different scales and angular modes of the cosmic density field. The resulting covariance matrix is therefore essential for properly combining these coefficients in the likelihood analysis.

\section{Results}\label{sec:results}

After validating the accuracy of the $R^{\rm wst}$ emulator, we now use it for Bayesian parameter inference. The validation tests in Section~\ref{sec:emulator} show that the emulator predictions are unbiased and that the emulator uncertainty is subdominant compared with the effective observational uncertainty. In this section, we investigate the constraining power of the WST $m$-mode ratio statistic and test its robustness against changes in tracer bias. We perform MCMC analyses on eight free parameters ($\Omega_b$, $\Omega_{m}$, $\sigma_8$, $n_s$, $w_0$, $w_a$, $H_0$, and $A_s$) in Sections~\ref{subsec:kun_inference_tests} and~\ref{subsec:rwst_tracer_bias}. In Section~\ref{subsec:comparison_2pcf_rwst}, we constrain only four parameters ($\Omega_{m}$, $\sigma_8$, $n_s$, and $w_0$), keeping the remaining four fixed.

\subsection{Inference tests on \texttt{Kun} cosmologies}
\label{subsec:kun_inference_tests}

Before applying the method to mock observational samples with different tracer biases, we first test the full inference pipeline on selected cosmologies from the \texttt{Kun} simulation suite. This provides a direct consistency check, since the true cosmological parameters of these simulations are known. We select four representative cosmologies, \texttt{C0000, C0002, C0004}, and \texttt{C0008}, and perform MCMC analyses using the \texttt{emcee} \cite{foreman-mackey_emcee_2013} sampler.

For each test cosmology, we use the $R^{\rm wst}$ data vector measured from the $M^A_{\rm cut}$ halo sample as the mock data vector. The theoretical prediction at each MCMC step is obtained from the trained GPR emulator, with the covariance matrix described in Section~\ref{sec:covariance_likelihood}. We vary eight cosmological parameters and adopt uniform priors over the simulation parameter ranges. Specifically, we use $\Omega_b \in [0.04,0.06]$, $\Omega_{m} \in [0.24,0.40]$, $\sigma_8 \in [0.54,1.18]$, $n_s \in [0.92,1.00]$, $w_0 \in [-1.30, -0.70]$, $w_a \in [-0.50,0.50]$, $H_0 \in [60,80]~{\rm km}~{\rm s}^{-1}~{\rm Mpc}^{-1}$, and $10^9A_s \in [1.70,2.50]$.

\begin{figure*}
\centering
\includegraphics[scale=0.61]{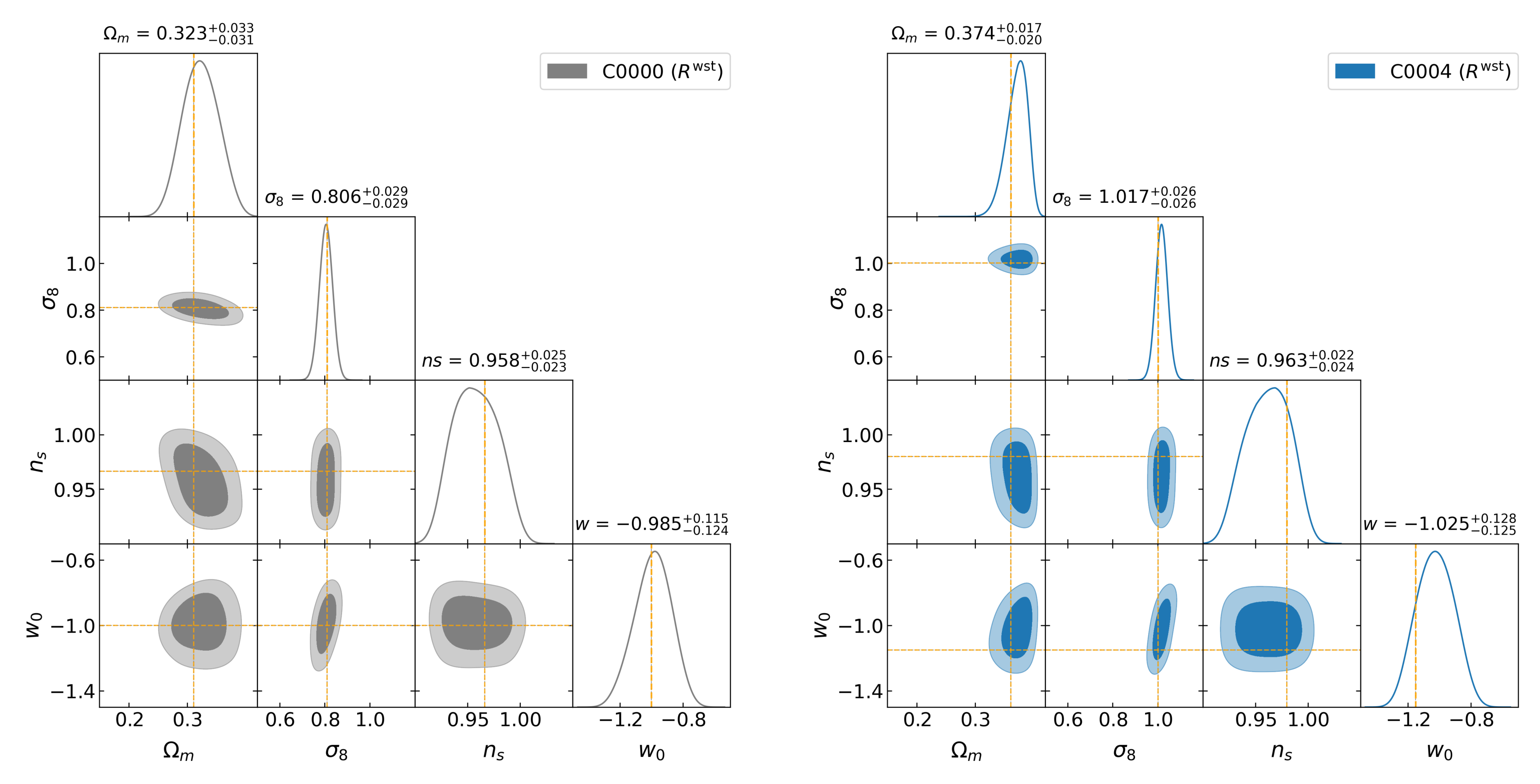}
\caption{
\label{fig:contour_combined_cosmic}
Marginalized $1\sigma$ and $2\sigma$ posterior contours obtained from two representative \texttt{Kun} cosmologies, \texttt{C0000}, and \texttt{C0004}, using the $R^{\rm wst}$ statistic with the $M^A_{\rm cut}$ selection. The MCMC analysis explores all eight cosmological parameters, while the figure highlights the constraints on the four parameters of primary interest: $\Omega_{m}$, $\sigma_8$, $n_s$, and $w_0$. The orange dashed lines indicate the true parameter values for each cosmology.
}
\end{figure*}

Figure~\ref{fig:contour_combined_cosmic} shows the marginalized posterior distributions for four representative parameters, $\Omega_{m}$, $\sigma_8$, $n_s$, and $w_0$, take \texttt{c0000} and \texttt{c0004} for examples. The orange dashed lines indicate the true parameter values of each selected cosmology. For all test cases, the true values are recovered within the marginalized $68\%$ credible regions. This confirms that the emulator-based likelihood pipeline is able to recover the input cosmology without significant bias.

Across the selected cosmologies, the $R^{\rm wst}$ statistic provides stable constraints on the displayed parameters. The typical marginalized uncertainties are about $ 0.025$ for $\Omega_{m}$, $0.027$ for $\sigma_8$, $0.023$ for $n_s$, and $0.123$ for $w_0$. These results demonstrate that the $R^{\rm wst}$ emulator can accurately recover cosmological parameters across a broad region of the \texttt{Kun} parameter space. This test therefore validates the inference pipeline before applying it to the tracer-bias robustness analysis below.

\subsection{Comparison between 2PCF and \bm{$R^{\rm wst}$}}
\label{subsec:comparison_2pcf_rwst}

To benchmark the constraining power of the WST $m$-mode ratio statistic, we compare it with a normalized 2PCF using the same inference pipeline. In this test, we focus on four cosmological parameters ($\Omega_{m}$, $\sigma_8$, $n_s$, and $w_0$), since the posterior distribution of the normalized 2PCF becomes poorly constrained when all eight parameters are varied. To facilitate a robust comparison of the constraining power, we fix the remaining four parameters ($\Omega_b$, $w_a$, $H_0$, and $A_s$) to their fiducial values. We adopt the same uniform priors as in the previous analysis.

For the 2PCF analysis, we use the normalized angularly averaged correlation function,
\begin{equation}
    \widehat{\xi}(s)=\frac{\xi(s)}{
    \int_{s_{\rm min}}^{s_{\rm max}} \xi(s')\,{\rm d}s'}\,.
\end{equation}
The angular average is performed over $\mu \in [0,0.96]$, where $\mu$ is the cosine of the angle between the pair separation and the line of sight. We use separation scales $s \in [15,85]~h^{-1}{\rm Mpc}$ with a bin width of $\Delta s = 10~h^{-1}{\rm Mpc}$.

\begin{figure}[htpb]
\centering
\includegraphics[scale=0.45]{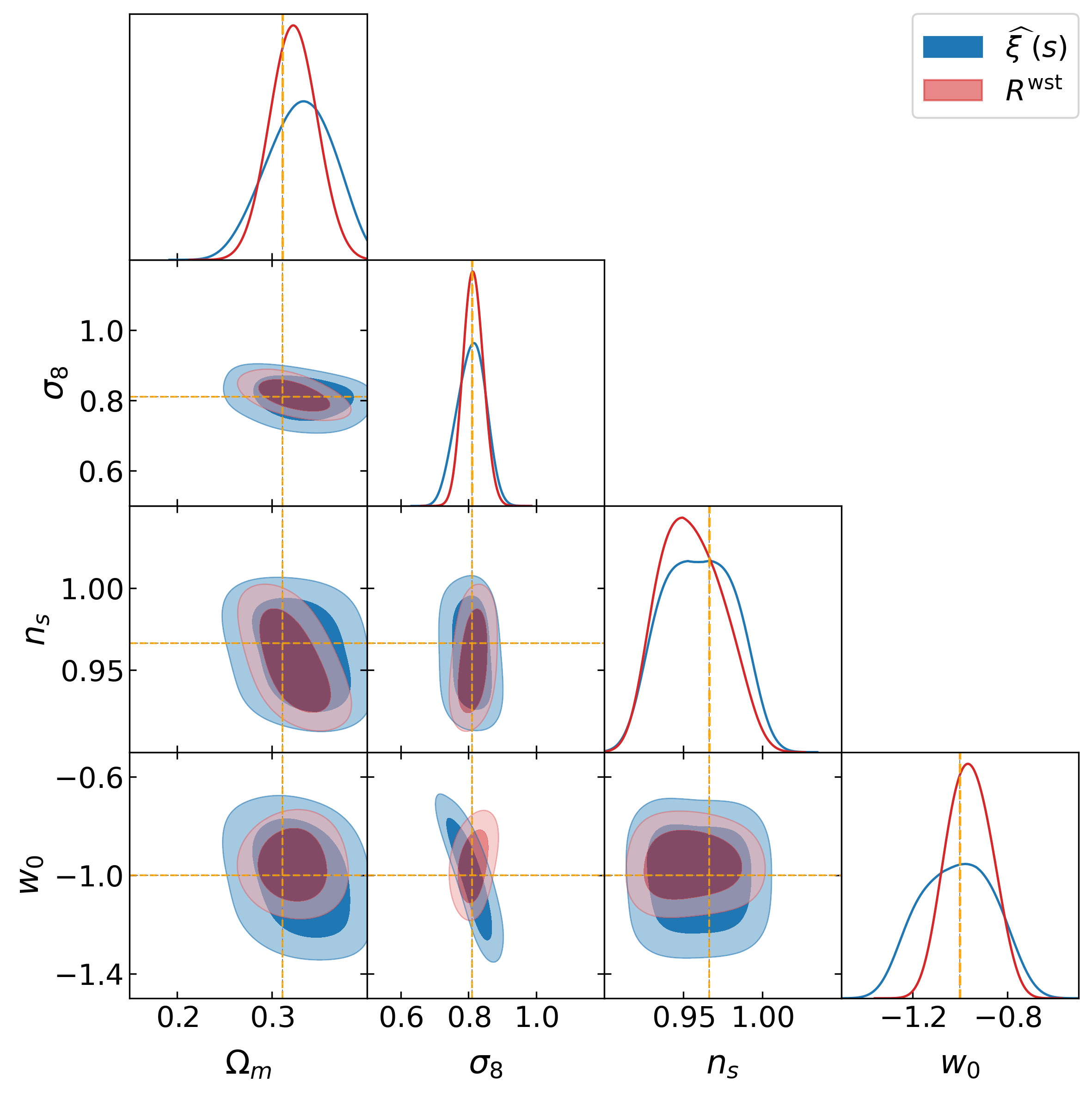}
\caption{
\label{fig:contour_vs_2pcf}
Marginalized $1\sigma$ and $2\sigma$ posterior distributions obtained from the fiducial \texttt{JiuTian} cosmology with the $M^A_{\rm cut}$ selection, comparing the constraints from $R^{\rm wst}$ and $\widehat{\xi}(s)$. The MCMC analysis is performed over four cosmological parameters, $\Omega_{m}$, $\sigma_8$, $n_s$, and $w_0$. The orange dashed lines indicate the true values of the \texttt{JiuTian} cosmology.
}
\end{figure}

Figure~\ref{fig:contour_vs_2pcf} compares the marginalized $1\sigma$ and $2\sigma$ posterior contours obtained from $R^{\rm wst}$ and $\widehat{\xi}(s)$ for the fiducial \texttt{JiuTian} cosmology with the $M^A_{\rm cut}$ selection. The corresponding posterior medians and marginalized $1\sigma$ uncertainties are listed in Table~\ref{tab:1}.

\begin{table}
\centering 
\caption{
Median values and marginalized $1\sigma$ uncertainties of $\Omega_{m}$, $\sigma_8$, $n_s$, and $w_0$ obtained from $\widehat{\xi}(s)$ and $R^{\rm wst}$, with the remaining cosmological parameters fixed to their fiducial values.
}
\label{tab:1} 

{
\renewcommand{\arraystretch}{1.35}
\begin{tabular}{lcccc}
\hline
\hline
 & $\Omega_{m}$ & $\sigma_8$ & $n_s$ & $w_0$   \\
\hline
$\widehat{\xi}(s)$
    & $0.331_{-0.038}^{+0.036}$ 
    & $0.811_{-0.046}^{+0.040}$ 
    & $0.960_{-0.026}^{+0.025}$ 
    & $-1.007_{-0.183}^{+0.165}$ \\
\hline
$R^{\rm wst}$
    & $0.323_{-0.024}^{+0.025}$ 
    & $0.813_{-0.029}^{+0.030}$ 
    & $0.954_{-0.021}^{+0.024}$ 
    & $-0.962_{-0.097}^{+0.101}$ \\
\hline
\hline
\end{tabular}
}

\end{table}

The $R^{\rm wst}$ statistic gives tighter marginalized constraints than $\widehat{\xi}(s)$ on $\Omega_{m}$, $\sigma_8$, $n_s$ and $w_0$. In particular, the uncertainty on $\Omega_m$ is reduced by $32\%$, from approximately $0.037$ for $\widehat{\xi}(s)$ to $0.025$ for $R^{\rm wst}$; the uncertainty on $\sigma_8$ is reduced by $30\%$, from approximately $0.043$ to $0.030$; and the uncertainty on $w_0$ is reduced by $43\%$, from approximately $0.174$ to $0.099$, indicating that the WST $m$-mode ratio captures non-Gaussian information beyond the two-point level.

To further quantify the joint constraining power, we compute the area of the $68\%$ credible contour in the $\Omega_{m}$--$\sigma_8$ plane. The contour area is about $0.0045$ for $R^{\rm wst}$ and $0.0101$ for $\widehat{\xi}(s)$. Thus, the $R^{\rm wst}$ constraint area is smaller by roughly a factor of two, indicating a weaker $\Omega_m$--$\sigma_8$ degeneracy and a stronger joint constraint than the standard 2PCF.

Overall, the WST $m$-mode ratio performs better than the 2PCF for four cosmological parameters, $\Omega_m$, $\sigma_8$, $n_s$ and $w_0$, providing stronger joint constraints on the full parameter set, especially through its improved sensitivity to $\Omega_m$, $\sigma_8$ and $w_0$.

\subsection{Robustness of \bm{$R^{\rm wst}$} to tracer bias}
\label{subsec:rwst_tracer_bias}

A key advantage of the WST $m$-mode ratio statistic is its robustness to tracer bias. Since $R^{\rm wst}$ is constructed from ratios between neighboring azimuthal WST modes, it is expected to be less sensitive to the overall amplitude modulation induced by tracer bias. This allows us to perform cosmological inference without explicitly introducing a halo-bias model, thereby reducing the risk of systematic errors associated with incomplete or inaccurate bias modeling.

To test this property, we perform MCMC analyses using mock data vectors measured from the \texttt{JiuTian} simulation under three halo-mass selections: $M^A_{\rm cut}$, $M^B_{\rm cut}$, and $M^C_{\rm cut}$. These samples span a wide range of effective tracer bias, as described in Section~\ref{subsec:analysis_datasets}. In all three cases, we keep both the emulator and the covariance matrix fixed to the fiducial $M^A_{\rm cut}$ setup. Therefore, any shift in the inferred parameters directly reflects the sensitivity of $R^{\rm wst}$ to changes in tracer bias. We vary eight cosmological parameters and adopt the same uniform priors over the simulation parameter ranges  as in the previous analysis.

\begin{figure}
\centering
\includegraphics[scale=0.45]{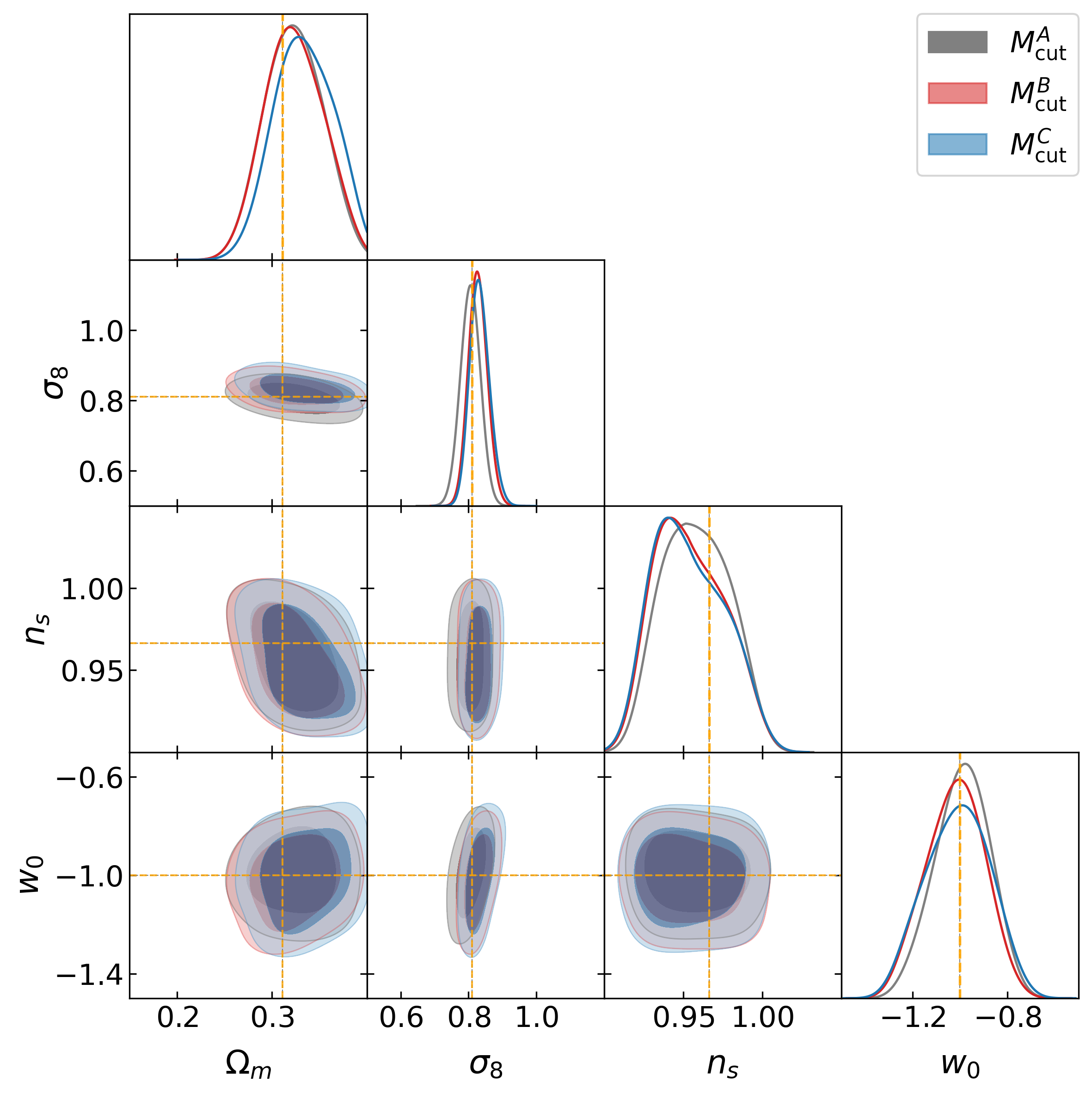}
\caption{
\label{fig:contour_vs_bias}
Marginalized $1\sigma$ and $2\sigma$ posterior distributions obtained from the fiducial \texttt{JiuTian} cosmology using the $R^{\rm wst}$ statistic under the $M^A_{\rm cut}$, $M^B_{\rm cut}$, and $M^C_{\rm cut}$ selections. The MCMC analysis varies eight free cosmological parameters, while the figure shows the constraints on $\Omega_{m}$, $\sigma_8$, $n_s$, and $w_0$. The orange dashed lines indicate the true parameter values of the \texttt{JiuTian} cosmology, as in Figure~\ref{fig:contour_vs_2pcf}.
}
\end{figure}

Figure~\ref{fig:contour_vs_bias} compares the marginalized posterior constraints on $\Omega_m$, $\sigma_8$, $n_s$, and $w_0$ for the three halo-mass selections. The corresponding posterior medians and marginalized $1\sigma$ uncertainties are listed in Table~\ref{tab:2}.

\begin{table}
\centering 
\caption{
Median values and marginalized $1\sigma$ uncertainties of $\Omega_{m}$, $\sigma_8$, $n_s$, and $w_0$ inferred from $R^{\rm wst}$ for the $M^A_{\rm cut}$, $M^B_{\rm cut}$, and $M^C_{\rm cut}$ halo selections, with all eight cosmological parameters varied in the MCMC analysis. The true values are $\Omega_{m}=0.3111$, $\sigma_8=0.8101$, $n_s=0.9665$, and $w_0=-1.0$.
}
\label{tab:2} 

{
\renewcommand{\arraystretch}{1.35}
\begin{tabular}{lcccc}
\hline
\hline
 & $\Omega_{m}$ & $\sigma_8$ & $n_s$ & $w_0$   \\
\hline
$M^A_{\rm cut}$
    & $0.323_{-0.031}^{+0.033}$ 
    & $0.806_{-0.029}^{+0.029}$ 
    & $0.958_{-0.023}^{+0.025}$ 
    & $-0.985_{-0.124}^{+0.115}$ \\
\hline
$M^B_{\rm cut}$
    & $0.322_{-0.030}^{+0.035}$ 
    & $0.827_{-0.026}^{+0.028}$ 
    & $0.952_{-0.022}^{+0.029}$ 
    & $-1.014_{-0.137}^{+0.121}$ \\
\hline
$M^C_{\rm cut}$
    & $0.333_{-0.031}^{+0.037}$ 
    & $0.831_{-0.027}^{+0.031}$ 
    & $0.951_{-0.022}^{+0.030}$ 
    & $-1.003_{-0.154}^{+0.139}$ \\
\hline
\hline
\end{tabular}
}

\end{table}

The recovered cosmological parameters remain broadly consistent with the true values for all three halo-mass selections. In particular, the inferred values of $\Omega_{m}$, $n_s$, and $w_0$ show random shifts between $M^A_{\rm cut}$, $M^B_{\rm cut}$, and $M^C_{\rm cut}$, and these shifts are well within their corresponding $1\sigma$ uncertainties. This indicates that these parameters are largely insensitive to the tracer-bias variation tested here.

For $\sigma_8$, we observe a mild upward shift as the selected halo mass range is moved toward lower masses. The difference between $M^A_{\rm cut}$ and $M^C_{\rm cut}$ is approximately at the $1\sigma$ level. This residual trend is physically expected because tracer bias and clustering amplitude are partially degenerate. Lower-mass halo samples have a lower effective bias and therefore a lower clustering amplitude. When the emulator and covariance are kept fixed to the fiducial $M^A_{\rm cut}$ setup, this change can be partially absorbed by a higher inferred value of $\sigma_8$.

Overall, these results show that $R^{\rm wst}$ is robust to substantial variations in tracer bias. Although a weak residual sensitivity remains for $\sigma_8$, the inferred parameters are stable at the level of the statistical uncertainties. This supports the use of the WST $m$-mode ratio statistic as a bias-robust summary statistic for cosmological inference.

\section{Conclusion}
\label{sec:conclusion}

The WST $m$-mode ratio statistic, $R^{\rm wst}$, is a summary statistic constructed from the wavelet scattering transform. The WST provides a multi-scale and nonlinear representation of the density field, making it sensitive to both Gaussian and non-Gaussian information. In our previous work~\citep{jiang_new_2025}, we showed that the ratio construction of $R^{\rm wst}$ can largely suppress the dependence on tracer bias while retaining cosmological sensitivity. This makes it a promising alternative to conventional summary statistics that usually require explicit bias modeling.

In this work, we developed a simulation-based inference framework to quantify the cosmological constraining power and tracer-bias robustness of $R^{\rm wst}$. We used the \texttt{Kun} simulation suite to train Gaussian Process Regression emulators for the $R^{\rm wst}$ data vector, allowing rapid predictions of the statistic as a function of cosmological parameters. The large-volume \texttt{JiuTian} simulation was used to estimate the covariance matrix and to construct mock observational data vectors with different halo-mass selections. This setup allowed us to test both the accuracy of the emulator and the stability of the inferred cosmological parameters.

We first validated the emulator using leave-one-out cross-validation. The emulator achieves high interpolation accuracy across the \texttt{Kun} cosmological parameter space, with relative errors below the percent level for all $R^{\rm wst}$ components. The emulator error is also generally subdominant compared with the effective observational uncertainty estimated from \texttt{JiuTian} subvolumes. These tests confirm that the emulator is sufficiently accurate for the subsequent MCMC parameter inference.

We then tested the full inference pipeline on selected \texttt{Kun} cosmologies with known input parameters. The inferred parameters are consistent with the true cosmological values within the marginalized $1\sigma$ credible regions, demonstrating that the emulator-based likelihood analysis is unbiased across a broad range of cosmologies. This result provides an important consistency check before applying the statistic to tracer-bias robustness tests.

Compared with the standard two-point correlation function, $R^{\rm wst}$ provides stronger joint constraints on cosmological parameters. The $R^{\rm wst}$ statistic gives a tighter marginalized constraint than the 2PCF on $\Omega_{m}$, $\sigma_8$, $n_s$ and $w_0$. Especially in the $\Omega_{m}$--$\sigma_8$ plane, the area of the $68\%$ credible contour is reduced from $0.0101$ for the 2PCF to $0.0045$ for $R^{\rm wst}$. This improvement indicates that $R^{\rm wst}$ captures additional non-Gaussian information beyond the two-point level.

We further examined the robustness of $R^{\rm wst}$ against tracer bias by applying the same emulator and covariance matrix, both constructed using the fiducial $M^A_{\rm cut}$ selection, to mock data vectors generated from $M^A_{\rm cut}$, $M^B_{\rm cut}$, and $M^C_{\rm cut}$. The inferred values of $\Omega_{m}$, $n_s$, and $w_0$ remain stable across the three halo-mass selections, with shifts much smaller than their corresponding $1\sigma$ uncertainties. This demonstrates that these parameters are largely insensitive to the tested changes in tracer bias.

For $\sigma_8$, we observe a mild upward shift as the selected halo mass range is moved toward lower masses. This residual trend is physically expected, because tracer bias and the clustering amplitude are partially degenerate. A lower-mass halo sample has a lower effective bias and therefore a lower clustering amplitude, which can be partly compensated by a higher inferred value of $\sigma_8$ when the emulator is fixed to the fiducial $M^A_{\rm cut}$ selection. Importantly, even for the extreme $M^C_{\rm cut}$ case, this shift remains at approximately the $1\sigma$ level. Therefore, the residual sensitivity does not significantly compromise the robustness of $R^{\rm wst}$ for cosmological inference.

Overall, our results show that the combination of $R^{\rm wst}$ and high-density apodization provides a powerful and bias-robust framework for extracting cosmological information from the nonlinear density field. The statistic improves joint parameter constraints relative to the standard 2PCF and remains stable under substantial changes in halo-mass selection. Since $R^{\rm wst}$ reduces the need for explicit tracer-bias modeling, it can help mitigate one of the major sources of systematic uncertainty in galaxy clustering analyses.

There are several directions for future work. First, the present analysis focuses on halo catalogs at $z=0.5$ in simulation boxes. A natural next step is to extend the test to multiple redshifts and more realistic galaxy samples generated with halo occupation or semi-analytic models. Second, observational effects such as survey geometry, masks, redshift-space distortions, selection functions, and shot noise should be included to assess the performance of $R^{\rm wst}$ in survey-like conditions. Finally, it would be valuable to combine $R^{\rm wst}$ with conventional statistics, such as the 2PCF or power spectrum, to determine how much complementary information can be extracted in joint analyses.

In summary, this work establishes $R^{\rm wst}$ as a promising non-Gaussian summary statistic for next-generation galaxy surveys. Its ability to improve cosmological constraints while remaining robust to tracer-bias variations makes it a useful tool for precision cosmology with Stage-IV surveys.

\begin{acknowledgements}
This work is supported by National SKA Program of China (2025SKA0160100), National Science Foundation of China (12473097), the China Manned Space Project with No. CMS-CSST-2021 (A02, A03, B01),  Guangdong Basic and Applied Basic Research Foundation (2024A1515012309), the National Natural Science Foundation of China (12373005).
\end{acknowledgements}
\bibliography{apssamp}{}

\end{document}